\documentclass[aip,jcp,reprint,floatfix,eqsecnum]{revtex4-1}

\usepackage{bm}

\usepackage{amssymb}

\usepackage{amsmath}

\usepackage{graphicx}      
\usepackage{dcolumn}       

\usepackage[usenames]{color}

\newcommand{\beq}{\begin{equation}}
\newcommand{\eeq}{\end{equation}}
\newcommand\beqa{\begin{eqnarray}}
\newcommand\eeqa{\end{eqnarray}}
\newcommand{\nn}{\nonumber\\}

\newcommand{\ed}{\end{document}}

\begin{document}


\title{On the relation between virial coefficients and the
close-packing of hard disks and hard spheres}



\author{Miguel \'Angel G. Maestre}
 \email{maestre@unex.es}

\author{Andr\'es Santos}
 \email{andres@unex.es}
 \homepage{www.unex.es/eweb/fisteor/andres/}
 \affiliation{Departamento de F\'{\i}sica, Universidad de Extremadura, Badajoz
06071, Spain}

\author{Miguel Robles}
\email{mrp@cie.unam.mx}
\homepage{http://xml.cie.unam.mx/xml/tc/ft/mrp/}

\author{Mariano \surname{L\'opez de Haro}}
\email{malopez@servidor.unam.mx}
\homepage{http://xml.cie.unam.mx/xml/tc/ft/mlh/}
\affiliation{Centro
de Investigaci\'on en Energ\'{\i}a, Universidad Nacional Aut\'onoma
de M\'exico (U.N.A.M.), Temixco, Morelos 62580, M{e}xico}

\date{\today}
\begin{abstract}
The question of whether the known virial coefficients are enough to determine
the packing fraction $\eta_\infty$ at which the fluid equation of state of a
hard-sphere fluid diverges is addressed. It is found that the information
derived from the direct Pad\'e approximants to the compressibility factor
constructed with the virial coefficients is inconclusive. An alternative
approach is proposed which makes use of the same virial coefficients and of the
equation of state in a form where the packing fraction is explicitly given as a
function of the pressure. The results of this approach both for hard-disk and
hard-sphere fluids, which can straightforwardly accommodate higher virial
coefficients when available, lends support to the conjecture that $\eta_\infty$
is equal to the maximum packing fraction corresponding to an ordered
crystalline structure.
\end{abstract}

\maketitle
\section{Introduction}
\label{intro}
The virial expansion of the equation of state {of a fluid} is an expansion in
powers of (usually) the number density $\rho$ that reads
\beq
\label{virial1}
Z(\rho,T)=1+\sum_{j=2}^\infty B_j(T) \rho^{j-1},
\eeq
where {$B_j$ are the virial
coefficients and} $Z\equiv p/\rho k_B T$ is the compressibility factor, {with}
$p$,
$k_B$, and $T$ being the pressure, the Boltzmann constant, and the
absolute temperature, respectively. This expansion was originally introduced by
Thiesen in 1885 (Ref.\ \onlinecite{T1885} in connection with the equation of state of fluids
at low densities. Soon afterwards, and apparently independently, it was used
by
Kamerlingh Onnes\cite{KO901} (who named the coefficients in the expansion as
virial coefficients), in order to provide a
mathematical representation of experimental results. The virial series was later
proven
to arise naturally in rigorous derivations in statistical
mechanics\cite{MM40} whereby the $B_j$ turn out to be related to
intermolecular interactions and are in general functions of
temperature.

In the case of hard-core systems such as hard disks or
hard spheres,{\cite{M08}} the virial  coefficients \emph{do not} depend on
temperature. In particular,
the value of the second virial coefficient for hard spheres {of diameter
$\sigma$} in $d$
dimensions is {$B_2= 2^{d-1} v_d\sigma^d$, where
$v_d=(\pi/4)^{d/2}/\Gamma(1+d/2)$ is the volume of a $d$-dimensional hard
sphere of unit diameter},
 a result first
derived for three-dimensional hard spheres ($d=3$) by van der
Waals.\cite{vdW899} Analytical expressions for $B_3$ and $B_4$ are
also available in the
literature\cite{J896,B896,vL899,B899,T36,R64,H64,LB82,BC87,CM04a,L05}
but higher virial coefficients must be computed numerically and,
since this represents a non trivial task, up to now only values up
to the tenth virial coefficient have been
reported.\cite{MRRT53,RR54,RH64a,RH64b,R65,RH67,KH68,K76,K77,K82a,K82b,vRT92,vR93,VYM02,CM04b,LKM05,CM05,KR06,CM06,BCW08}

The virial expansion for $d$-dimensional hard-sphere
systems is often expressed in terms of the packing fraction $\eta$
defined as
\beq
\label{packing}
\eta=v_d \rho \sigma^d.
 \eeq
Hence, for these systems the
virial expansion of the compressibility factor is given by
\beq
\label{virial2}
Z(\eta)=1+\sum_{j=2}^\infty b_j \eta^{j-1},
\eeq
where the (reduced) virial coefficients $b_j\equiv B_j/(v_d
\sigma^d)^{j-1}$  are pure numbers.
{For one-dimensional hard rods, $b_j=1$ and so $Z(\eta)=1/(1-\eta)$.} The
presently known values\cite{CM06} of the virial coefficients  coefficients for
hard disks ($d=2$) and hard spheres ($d=3$) are given in {Table \ref{tab1}}.

\begin{table}
\caption{Known virial coefficients\cite{CM06} for a hard-disk fluid ($d=2$) and
a hard-sphere fluid ($d=3$).
\label{tab1}}
\begin{ruledtabular}
\begin{tabular}{ccc}
$j$ &\mbox{$b_j$ ($d=2$)}&\mbox{$b_j$ ($d=3$)}\\
\hline
${2} $ & $2$&$4$\\
$ {3} $ & $3.12801775\ldots$&$10$\\
$ {4} $ & $4.25785445\ldots$&$18.36476838\ldots$\\
$ {5} $ & $5.3368966(2)$&$28.22451(26)$\\
${6} $ & $6.36296(13)$&$39.81515(93)$\\
$ {7} $ & $7.35186(28)$&$53.3444(37)$\\
$ {8} $ & $8.31910(44)$&$68.538(18)$\\
$ {9} $ & $9.27215(90)$&$85.813(85)$\\
$ {10} $ & $10.2163(41)$&$105.78(39)$\\
\end{tabular}
\end{ruledtabular}
\end{table}

There are {a number of} controversial {and open} issues related to the virial
expansion
of hard-disk and hard-sphere fluids.{\cite{M10}} To begin with, even in the
case that many more virial coefficients for these systems were known, the
truncated virial series for the corresponding compressibility factors would not
be useful in principle for packing fractions higher than the one corresponding
to the radius of convergence  $\eta_{\text{conv}}$ of the whole series. Such
{a} radius of convergence is determined by the modulus of the singularity of
{$Z(\eta)$} closest to the origin in the complex plane. While its actual value
is not known, lower bounds are available\cite{LP64,FPS07} and existing evidence
suggests that it derives from a singularity located on the real negative
axis.\cite{CM06,SH09} In fact, one of the major reasons for trying to evaluate
higher order virial coefficients is precisely the determination of the value of
$\eta_{\text{conv}}$ and of the nature of the singularity giving rise to it.
One should add that, although the virial series diverges for $\eta
>\eta_{\text{conv}}$, given the fact that $\eta_{\text{conv}}$ seems to be
located outside the positive real axis, one would expect that the
compressibility factor would still be well defined for $\eta
\geq\eta_{\text{conv}}$, at least in a certain range.

Coming back to the controversial {and open} issues, even the character of
the
virial expansion (either alternating or not) is still unknown. So far, all
the available virial coefficients for these systems are positive,
but results from higher dimensions suggest that this feature might
not be true for higher virial coefficients.\cite{CM06} Finally, the
evidence coming from approximate equations of state obtained {through}
the knowledge of the limited number of virial coefficients via
various series acceleration methods, such as Pad\'e or Levin
approximants, indicates that the freezing transition observed in
computer simulations does not show up {as a singularity} in these equations of
state.\cite{AFLlR84}
{As a matter of fact, while it is quite plausible that  $Z(\eta)$ presents a
singularity at the freezing density $\eta_f$,\cite{GJ80,J88,BET80} the
virial coefficients (or even their asymptotic behavior) do not seem to yield
any information concerning the freezing transition at
$\eta_f$.\cite{GJ80} This might be related to the fact that
$\lim_{\eta\to\eta_f^-}Z(\eta)=\text{finite}$.}

Since the compressibility factor of hard-disk and hard-sphere fluids both for
the stable and metastable fluid phases is a monotonically increasing function
of the packing fraction,{\cite{M10}} one may reasonably wonder at which
packing fraction $\eta=\eta_\infty$ the {analytical continuation of the}
compressibility factor diverges, namely one wants to find the value of
$\eta_\infty$ such that
\beq
\lim_{\eta\to\eta_{\infty}}Z(\eta)=\infty.
\label{16}
\eeq
Clearly, $\eta_\infty$ may not be bigger than the maximum packing fraction
$\eta_{\text{max}}$ that is geometrically possible  (the so-called Kepler's
problem).  Therefore, one must have
\beq
\eta_{\text{conv}}\leq\eta_\infty\leq \eta_{\text{max}}.
\label{17}
\eeq
The value of $\eta_{\text{max}}$, at least for not too high dimensionalities
$d$,  corresponds to an ordered crystalline structure. In Table \ref{tab2} we
provide the values of $\eta_{\text{max}}$ for $d=2$--$8$.

More than a decade ago, by studying the singularities of {Pad\'e approximants
constructed from} the virial
series for hard disks and hard spheres, Sanchez\cite{S94} came to
the conclusion that such singularities {were} related to
crystalline close-packing in these systems. Other
authors\cite{W76,A76,BL79,DS82,AFLlPRR83,HvD84,GW88,SHY95,WKV96,KV97,NAM00,GV01,W02,PV05,MMD06}
have also conjectured that
\beq
\eta_\infty= \eta_{\text{max}}.
\label{18}
\eeq
{It is interesting to note that this conjecture was already suggested by
Korteweg\cite{K92} and Boltzmann\cite{B98} in the late 1800s.}

On the other hand, the conjecture \eqref{18} has not been free from
criticism\cite{GJ80}  {and some authors\cite{F72,AFLlPR82,MA86,SSM88,H97} have
conjectured that $\eta_\infty=\eta_{\text{rcp}}$, where  $\eta_{\text{rcp}}$
(approximately equal to $0.82$ and $0.64$ for hard disks and hard spheres,
respectively\cite{B83}) is the random close-packing fraction. For a
{thorough} account of proposed equations of state, including those enforcing
  $\eta_\infty= \eta_{\text{max}}$ or $\eta_\infty=\eta_{\text{rcp}}$, the
reader is referred to Ref.\ \onlinecite{MGPC08}.}

In his original study, Sanchez used a
Pad\'e analysis with the then most recent available values of the
first eight virial coefficients given by Janse van
Rensburg.\cite{vR93} More recently, Sanchez and Lee\cite{SL09} found that, using
the ninth and tenth virial coefficients\cite{CM06} for a hard-sphere fluid, the
corresponding Pad\'e approximants remained finite at $\eta=\eta_{\text{max}}$.
However, taking into account such coefficients, they constructed a new
approximate equation of state that diverges at that packing fraction. In this
regard, the question arises as to whether using the available information on
virial coefficients one can reach a more definite conclusion. More generally,
one could ask whether there may be a systematic method to improve the
estimation of $\eta_\infty$ as more virial coefficients become available.

The paper is organized as follows. In Sec.\ \ref{sec2} we perform an analysis of
the singularities of the compressibility factors obtained from the different
Pad\'e approximants constructed from the known virial coefficients of the
{hard-disk and hard-sphere fluids}. This is followed in Sec.\ \ref{sec3} by
the introduction of inverse representations of the virial series and their
connection with the computation of $\eta_\infty$, {the results being presented
in Sec.\ \ref{sec4}}. The paper is closed in Sec.\ \ref{sec5} with some
concluding remarks.

\begin{table}
\caption{Values of $\eta_{\text{max}}$ for $d=2$--$8$.\protect\cite{lattices}
\label{tab2}}
\begin{ruledtabular}
\begin{tabular} {ccc}
$d$ & Exact & Numerical\\
\hline
$ 2 $ &  $\frac{1}{6} \pi \sqrt{3}$  & $0.9069$\\
$ 3 $ &  $\frac{1}{6} \pi \sqrt{2}$ & $0.7405$\\
$ 4 $ & $\frac{1}{16} \pi^{2}$ & $0.6168$\\
$ 5 $ & $\frac{1}{30} \pi^{2} \sqrt{2}$ & $0.4652$\\
$ 6 $ & $\frac{1}{144} \pi^{3} \sqrt{3}$ & $0.3729$\\
$ 7 $ & $\frac{1}{105} \pi^{3}$ & $0.2953$\\
$ 8 $ & $\frac{1}{384} \pi^{4}$ & $0.2537$\\
\end{tabular}
\end{ruledtabular}
\end{table}

\section{Pad\'e approximants and singularities in the compressibility factor}
\label{sec2}

In general, given the series
\beq
S(z)= \sum_{j=0}^\infty a_{j} z^{j},
\label{2.1}
\eeq
the notion behind a Pad\'e approximant is the replacement of the series by a
ratio of two polynomials, namely
\begin{equation}
S(z)\approx P_{N}^{M}(z)=\frac{\sum_{j=0}^{M} \alpha_{j}z^{j}}{\sum_{j=0}^{N}
\beta_{j}z^{j}},
\label{2.1b}
\end{equation}
where, without loss of generality, one may take $\beta_{0} =1 $. The remaining $
N+M+1 $ coefficients are chosen in such a way that the Taylor series expansion
of $ P_{N}^{M} (z) $ exactly yields the first $ N+M+1 $ terms of the power
series $S(z)$. On the other hand, the difference $N-M$ of the degrees of the
two polynomials may be taken at will. A more detailed treatment of Pad\'e
approximants may be found in Ref. \onlinecite{BO78}.

In the usual application of Pad\'e approximants for the equation of state of
hard-sphere fluids,  the compressibility factor $Z$  is approximated by
\begin{equation}
Z(\eta)\approx P_{N}^{M}(\eta)=\frac{1+\sum_{j=1}^{M}
\alpha_{j}\eta^{j}}{1+\sum_{j=1}^{N} \beta_{j}\eta^{j}},
\label{2.2}
\end{equation}
where, since $\lim_{\eta\to 0}Z(\eta)=1$, one sets $\alpha_0=1$. The other
$N+M$ coefficients $\{\alpha_{j}, j=1,\ldots,M\}$ and $\{\beta_j,
j=1,\ldots,N\}$ are determined with the values of the virial coefficients
$\{b_j, j=2,\ldots N+M+1\}$. For a given value of $g\equiv N+M$, there are in
principle $g$ independent approximants (corresponding to $N=1,\ldots,g$),
because the case $(N,M)=(0,g)$ is not a Pad\'e approximant but rather a
truncated series. We will use the term `Pad\'e degree' for $g=N+M$. Given the
known values of $b_j$ (see Table \ref{tab1}), the highest value of the Pad\'e
degree is $g=9$.
{The Pad\'e approximants \eqref{2.2} are not only used to get the equation of
state but  also  to predict values of unknown virial
coefficients.\cite{GB08,HY09}}

Both the radius of convergence $\eta_\text{conv}$ of the virial series and the
value of $\eta_\infty$ when $Z$ is {approximated} by \eqref{2.2} are determined
by the nature of the singularities of the approximant.
{The modulus of the complex root of $1+\sum_{j=1}^{N} \beta_{j}\eta^{j}$ closest
to the origin gives $\eta_\text{conv}$, while $\eta_\infty$ is identified with
the smallest real positive root.}

\begin{figure}
\includegraphics[width=.95\columnwidth]{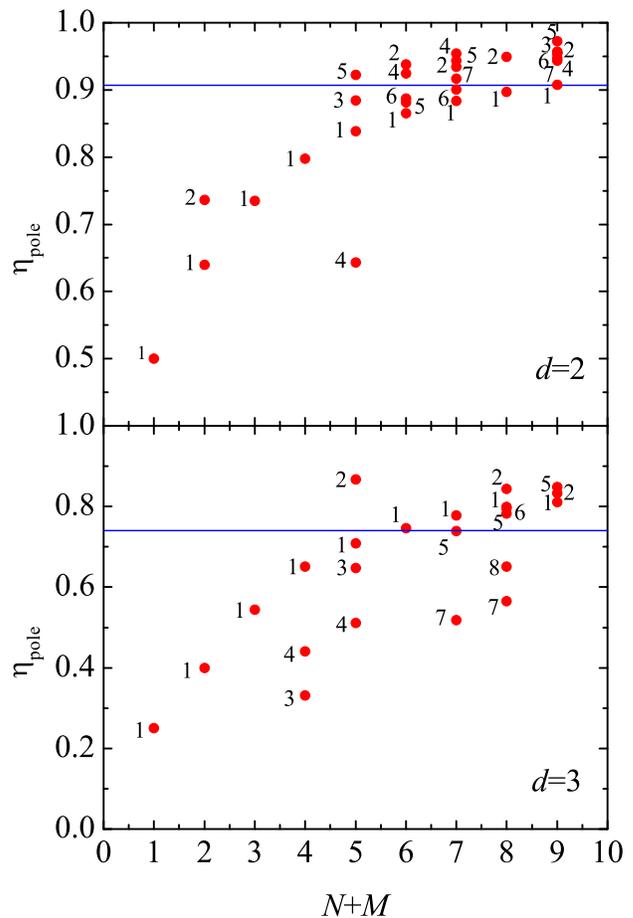}
\caption{Value of the real and positive pole closest to the origin for the
different Pad\'e approximants {of the form \protect\eqref{2.2}} as a function
of the Pad\'e degree $g=N+M$. The number placed next to each circle indicates
{the degree $N$ of the polynomial  in the denominator} of the corresponding
Pad\'e approximant. The horizontal lines correspond to
$\eta=\eta_{\text{max}}$.\label{fig1}}
\end{figure}

Due to our interest in obtaining the value of $\eta_\infty$, we have carried out
a systematic analysis of the location of the \emph{real and positive} pole of
$P_{N}^{M}(\eta)$ closest to the origin. For each Pad\'e degree $g\leq 9$ we
have considered all the associated $g$ Pad\'e approximants, namely a total of
$\sum_{g=1}^9 g=45$ approximants, and located their poles.  In some instances,
the corresponding approximant either has no real and positive poles or these
are located beyond $\eta=1$. In the rest of the cases, if more than one real
and positive pole existed in the range $0<\eta<1$, we have focused on the
smallest one of them. {Moreover, we have discarded a pole when it practically
coincided with a zero of the approximant.} The results of this analysis are
shown in Fig.\ \ref{fig1}.

As can be seen in this figure, {only $29$ ($d=2$) or $23$ ($d=3$) out of the 45
approximants possess} a real and positive pole in the range $0<\eta<1$.
{The Pad\'e approximants with $N=1$  have a real pole at
$\eta_\text{pole}=b_{g}/b_{g+1}$ and this is  the only pole for a few values of
$g$.} Also worth pointing out is the fact that the real and positive
singularity closest to the origin exhibits an important dispersion for a given
Pad\'e degree $g$, something particularly noticeable for $g=8$ {and $d=3$.
Moreover,  for $g\geq 5$} the value of the poles may even be higher than
$\eta_{\text{max}}$, which is clearly unphysical.

\begin{figure}
\includegraphics[width=.95\columnwidth]{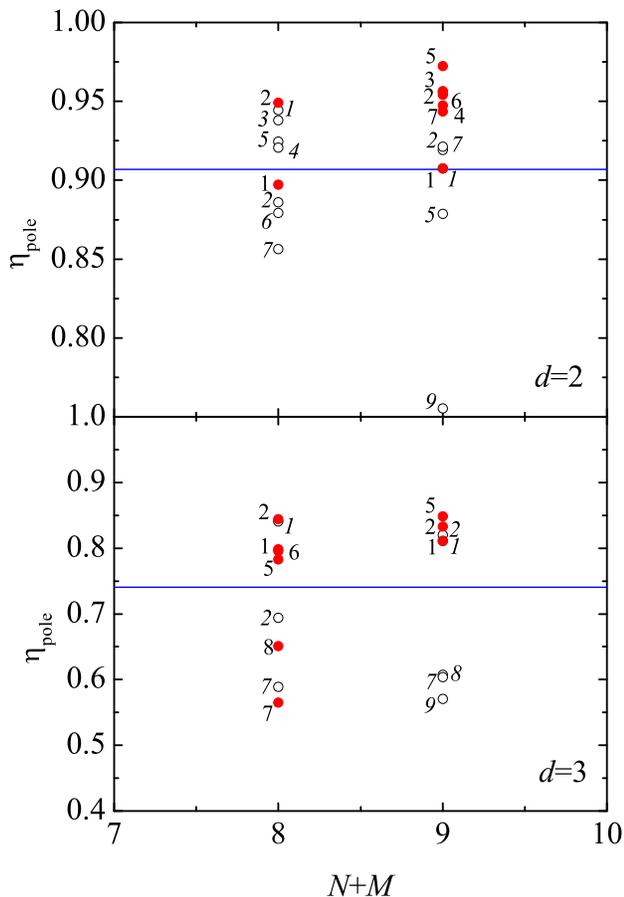}
\caption{Value of the real and positive pole closest to the origin for $g=8$ and
 $g=9$. Filled circles correspond to the values obtained from the use of the
coefficients  $b_9$ and $b_{10}$ given in Table \protect\ref{tab1}, while the
open circles have been obtained using the values {$(b_{9},b_{10})= (8.80,9.71)$
for $d=2$ and $(b_{9},b_{10})= (81.52,100.5)$ for $d=3$}.  The number placed
next to each circle (in italics for open circles) indicates {the degree $N$ of
the polynomial  in the denominator} of the corresponding Pad\'e approximant.
The horizontal lines correspond to  $\eta_{\text{max}}$.\label{fig2}}
\end{figure}

In order to check on the robustness of the method of analysis, we have carried
out the same analysis for the cases of $g=8$ and $g=9$, this time varying the
values of the two last known virial coefficients by $5\%$. {More specifically,
instead of the values of $b_9$ and $b_{10}$ given in Table \ref{tab1}, we have
taken  $(b_{9},b_{10})= (8.80,9.71)$ for $d=2$ and $(b_{9},b_{10})=
(81.52,100.5)$ for $d=3$}.
The consequence of such replacement is an important variation in the character
and location of the poles, as shown in Fig.\ \ref{fig2}, including the
appearance of new poles and the disappearance of earlier ones. {An exception is
the pole $\eta_\text{pole}=b_{9}/b_{10}$ corresponding to $N=1$ and $g=9$,
which obviously is not affected by a common factor multiplying both $b_9$ and
$b_{10}$.} Given this generally great sensitivity of the poles on the values of
the virial coefficients, it does not seem wise to rely on the previous analysis
in order to get good estimates of $\eta_\infty$. Instead, in Sec.\ \ref{sec3}
we will follow a different route.

\section{Inverse representations of the virial series}
\label{sec3}
{We have seen that the use of the direct Pad\'e approximants of the compressibility factor $Z$ is not reliable for the purpose of determining $\eta_\infty$. This is perhaps not surprising since, as
McCoy\cite{M10} has pointed out in his recent book (and we quote), ``there is no reason to expect
that simple model equations of state which use the density as the independent
variable
will capture the true physics of hard particle systems (or indeed of any real
system).''
Therefore, an alternative approach is called for.} We begin by introducing an equivalent form of the virial series \eqref{virial2},
namely
\begin{equation}
\widetilde{p}(\eta) \equiv v_d \sigma^d \frac{p}{k_BT} = \eta +
\sum_{j=2}^{\infty} b_{j}  \eta^{j}.
\label{3.1}
\end{equation}
{Note that $\widetilde{p}=\eta Z$.}
Following an idea of Sanchez,\cite{S94} we may formally \emph{invert}  the
series in \eqref{virial2},  leading to
\begin{equation}
\eta(Z) =\sum_{j=1}^{\infty} e_{j}  (Z-1)^{j}.
\label{3.5}
\end{equation}
{Here we propose a similar inversion of  the series \eqref{3.1} in the form}
\begin{equation}
\eta(\widetilde{p}) = \widetilde{p} + \sum_{j=2}^{\infty} c_{j}
\widetilde{p}^{j}.
\label{3.4}
\end{equation}
The coefficients $\{c_j\}$ and $\{e_j\}$  are (nonlinear) combinations of the
virial coefficients $\{b_j\}$. In fact, the determination of $c_j$ and $e_j$
requires that one previously knows $j$ and $j+1$ virial coefficients,
respectively. Notice that, while going from  Eq.\ \eqref{virial2} to Eq.\
\eqref{3.1} or vice versa is a trivial task, the same is not true when
considering the inverse developments \eqref{3.5} and \eqref{3.4}. Further, the
condition \eqref{16} to obtain $\eta_\infty$ is equivalent to
\beq
\eta_\infty=\lim_{Z\to\infty}\eta(Z)=\lim_{\widetilde{p}\to\infty}\eta(\widetilde{p}).
\label{eta}
\eeq

The form \eqref{3.5} was the one that Sanchez used\cite{S94} to estimate the
value of $\eta_\infty$. However, the true \emph{thermodynamic} variable, along
with the density (hereby represented by $\eta$), is the pressure (hereby
represented by $\widetilde{p}$), and not the ratio pressure over density (given
by $Z$). Therefore, in our view the form \eqref{3.4} has a clearer physical
meaning than \eqref{3.5}.
{In this respect, it is interesting to note that  Hamad\cite{H97,HY00} proposed
for hard-disk and hard-sphere fluids an approximate equation of state of the
form
\beq
\eta(\widetilde{p})=\frac{\widetilde{p}}{1+b_2\widetilde{p}-k_1\widetilde{p}\ln\frac{1+k_2\widetilde{p}}{1+k_3
\widetilde{p}}},
\label{Hamad}
\eeq
with $k_1=(b_2^2-b_3)/(k_2-k_3)$ and where $k_2$ and $k_3$ were obtained by a
fitting method. According to Eq.\ \eqref{Hamad},
$\eta_\infty=[b_2-k_1\ln(k_2/k_3)]^{-1}$, which turns out to be close to the
respective random close-packing values.\cite{H97}}

There is an additional argument in favor of expressing the equation of state in
the form  $\eta=\eta(\widetilde{p})$. For \emph{one-dimensional} systems of
particles interacting through an \emph{arbitrary} potential $\varphi(r)$
(restricted to nearest neighbors) there is an \emph{exact}
statistical-mechanical solution.\cite{SZK53,CD98,HC04} In this solution, the
equation of state is given by\cite{FGMS09}
\beq
\rho(p,T)=\frac{\Omega_0(p/kT,T)}{\Omega_1(p/kT,T)},
\label{1D}
\eeq
where
\beq
 \Omega_s(x,T)\equiv \int_0^\infty dr\, r^s e^{-xr}e^{-\varphi(r)/kT}.
\label{1Db}
\eeq
Thus, at a given temperature, the density is an explicit function of the
pressure. In general (the hard-rod case is an exception), it is not possible to
invert Eq.\ \eqref{1D} to express the pressure as an explicit function of
density {or the density as an explicit function of $Z$}. The functional form
\eqref{1D} may be easily extended to the case of mixtures.\cite{BNS09}

\begin{figure}
\includegraphics[width=.95\columnwidth]{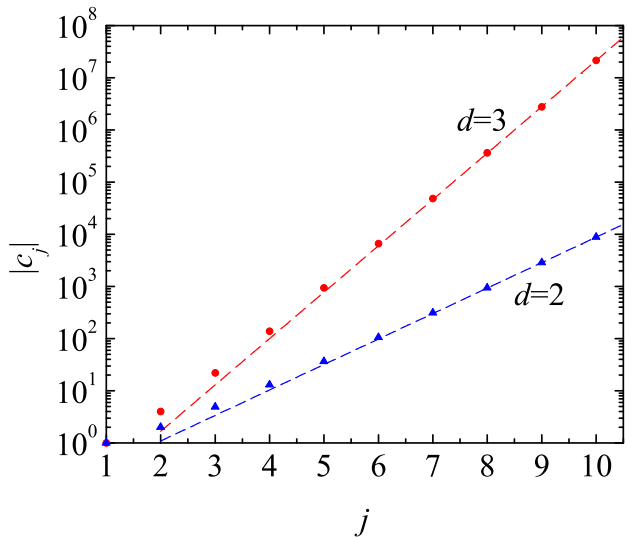}
\caption{{Semi-logarithmic plot of the absolute value of the inverse virial
coefficients $c_j$ for $j\leq 10$. The dashed lines represent the exponential
fit $|c_j|\approx c_9(|c_{10}|/c_9)^{j-9}$.}
\label{fig3}}
\end{figure}

{We have obtained the values of the inverse virial coefficients $c_j$ for $2\leq
j\leq 10$ and  $e_j$ for $1\leq j\leq 9$ from the knowledge of the first ten
virial coefficients listed in Table \ref{tab1}. Here we give some details about
the coefficients $c_j$. It is observed that they  alternate sign ($c_2=-b_2$
being negative) and their absolute values grow almost exponentially with $j$.
The results are graphically displayed in Fig.\ \ref{fig3}.}

{For further use, it is interesting to provide the expressions of the
coefficients $c_9$ and $c_{10}$ when $b_2$--$b_8$ are given by Table \ref{tab1}
but $b_9$ and $b_{10}$ remain free. For hard disks, one gets}
\beq
{c_9=2.86\times 10^3\left(1-3.49\times 10^{-4} b_9\right)},
\label{012D}
\eeq
\beq
{c_{10}=-8.97\times 10^3\left(1-2.45\times 10^{-3} b_9+1.11\times 10^{-4}
b_{10}\right).}
\label{022D}
\eeq
{Since $b_9$ and $b_{10}$ are of the order of $10$, it is clear that neither
$c_9$ nor $c_{10}$ are strongly affected by the precise values of $b_9$ and
$b_{10}$. This effect is even much more pronounced in the case of hard spheres,
where}
\beq
{c_9=2.78\times 10^6\left(1-3.60\times 10^{-7} b_9\right),}
\label{01}
\eeq
\beq
{c_{10}=-2.15\times 10^7\left(1-2.05\times 10^{-6} b_9+4.66\times 10^{-8}
b_{10}\right).}
\label{02}
\eeq
{Taking into account that both $b_9$ and $b_{10}$ are of order $10^2$, one sees
that the actual values of $c_9$ and $c_{10}$ are very weakly influenced by
$b_9$ and $b_{10}$. The same process can be repeated for $c_{11}$ and $c_{12}$,
this time using the values of $b_9$ and $b_{10}$ listed in Table \ref{tab1} but
leaving $b_{11}$ and $b_{12}$ free. The results for hard disks are}
\beq
{c_{11}=2.72\times 10^4\left(1-3.67\times 10^{-5} b_{11}\right),}
\label{032D}
\eeq
\beq
{c_{12}=-8.53\times 10^4\left(1-3.05\times 10^{-4} b_{11}+1.17\times 10^{-5}
b_{12}\right).}
\label{042D}
\eeq
{Analogously, for hard spheres one finds}
\beq
{c_{11}=1.68\times 10^8\left(1-5.96\times 10^{-9} b_{11}\right),}
\label{03}
\eeq
\beq
{c_{12}=-1.32\times 10^9\left(1-3.92\times 10^{-8} b_{11}+7.55\times 10^{-10}
b_{12}\right).}
\label{04}
\eeq
{Therefore, as $j$ increases, the values of $c_j$  become less and less
sensitive to the actual values of the virial coefficients $b_{j-1}$ and $b_j$,
this effect being much more important for hard spheres than for hard disks.
This implies that one could get good estimates of $c_{11}$ and $c_{12}$
(especially for hard spheres) even with poor estimates of the unknown virial
coefficients $b_{11}$ and $b_{12}$. We will come back to this point later.}

Once we have introduced the representations \eqref{3.5} and \eqref{3.4}, the
next step is to compute their corresponding Pad\'e approximants. These read
\begin{equation}
\eta(Z) \approx (Z-1) P_N^{N-1}(Z-1)
\label{3.8}
\end{equation}
and
\begin{equation}
\eta(\widetilde{p}) \approx \widetilde{p} P_N^{N-1}(\widetilde{p}).
\label{3.7}
\end{equation}
Now one can determine $\eta_\infty$ using Eq.\ \eqref{eta}. Since $\eta_\infty$
must be finite and different from zero, we must restrict ourselves {to
diagonal Pad\'e approximants, in contrast with what occurs with the direct
approximants \eqref{2.2}.} Therefore,
\beq
\eta_\infty(N)=\frac{\alpha_{N-1}}{\beta_N},
\label{eta2}
\eeq
where, following the notation of Eq.\ \eqref{2.1b}, $\alpha_{N-1}$ and $\beta_N$
are the coefficients of the terms of the highest degree in the numerator and
denominator of the approximant, respectively. In Eq.\ \eqref{eta2} the notation
$\eta_\infty(N)$  indicates that, in principle, the value so obtained may
depend on $N$. One would expect that the true value of $\eta_\infty$ would be
$\eta_\infty=\lim_{N\to\infty}\eta_\infty(N)$.

Since $b_1=1$, in the case of Eq.\ \eqref{3.8} we have to determine $2N$
coefficients using  $e_1$,\ldots $e_{2N}$ (or, equivalently, $b_2$,\ldots
$b_{2N+1}$). On the other hand, in order to use Eq.\ \eqref{3.7}, $2N-1$
{coefficients} must be determined from $c_2$,\ldots $c_{2N}$ (or,
alternatively, from $b_2$,\ldots $b_{2N}$). We recall that only ten virial
coefficients are presently known, so the highest $N$ that in principle may be
taken in connection with Eqs.\ \eqref{3.8} and \eqref{3.7} is $4$ and $5$,
respectively. This restriction may be removed at the expense of including
\emph{estimates} of the first few virial coefficients beyond the tenth.

\section{Results}
\label{sec4}

\begin{figure}
\includegraphics[width=.95\columnwidth]{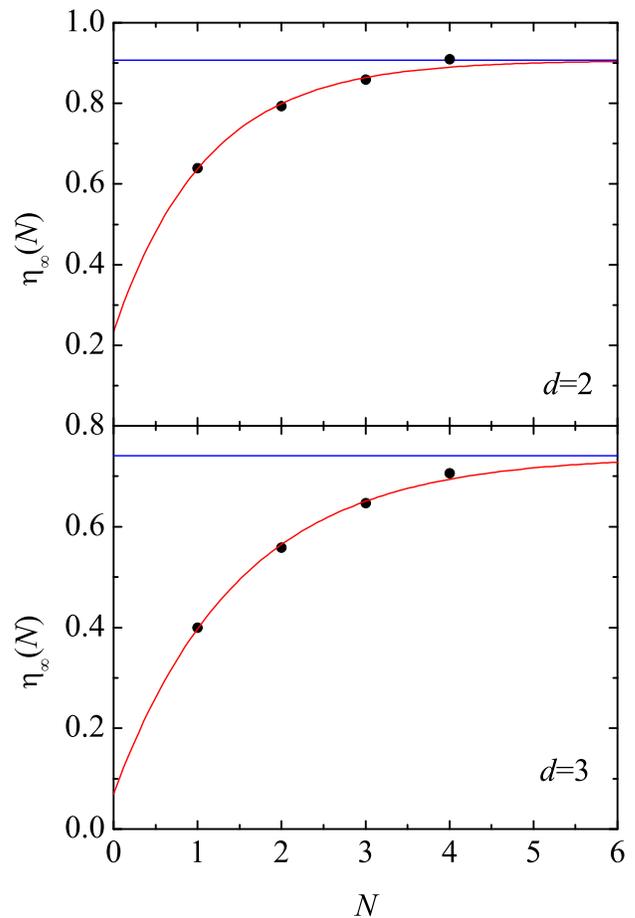}
\caption{Estimates of $\eta_\infty$ derived from Eq.\ \protect\eqref{3.8}. The
horizontal lines correspond to $\eta_{\text{max}}$. The curves correspond to
the fit  {$\eta_\infty(N)=\eta_{\text{max}} -0.67  e^{-0.92 N}$ ($d=2$) and
$\eta_\infty(N)=\eta_{\text{max}} -0.67  e^{-0.67 N}$ ($d=3$)}.
\label{fig4}}
\end{figure}

\begin{figure}
\includegraphics[width=.95\columnwidth]{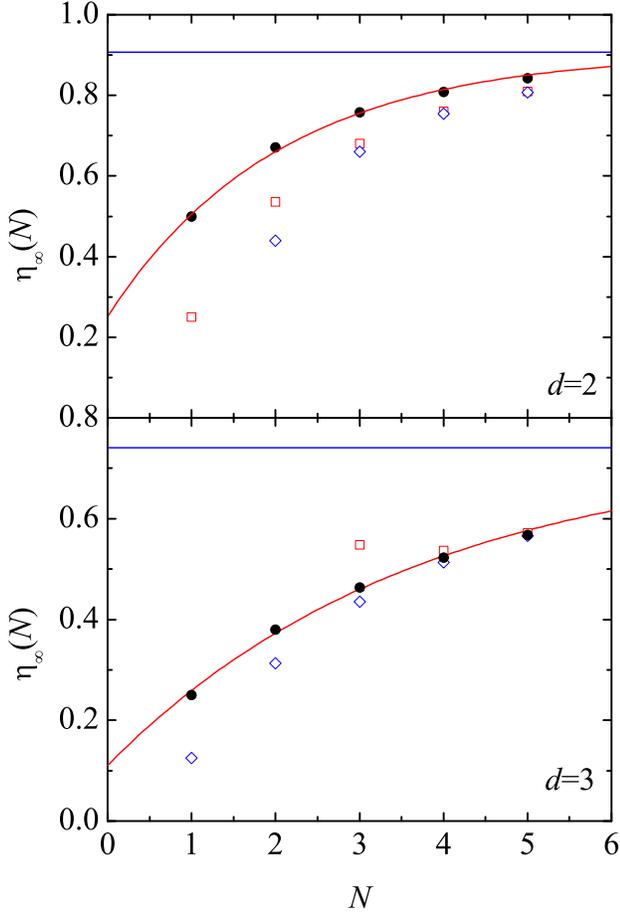}
\caption{Estimates of $\eta_\infty$ derived from Eq.\ \protect\eqref{3.7}. The
horizontal lines correspond to $\eta_{\text{max}}$. The curves correspond to
the fit   {$\eta_\infty(N)=\eta_{\text{max}} -0.65  e^{-0.49 N}$ ($d=2$) and
$\eta_\infty(N)=\eta_{\text{max}} -0.63  e^{-0.27 N}$ ($d=3$). The circles,
squares, and diamonds correspond to taking for $b_{2N}$ its correct value given
in Table \ref{tab1}, zero, and twice the correct value, respectively}.
\label{fig5}}
\end{figure}

The results for $\eta_\infty(N)$, derived from the different Pad\'e approximants
as given by Eq.\ \eqref{3.8} (with $N=1,2,3,4$) and the use of Eq.\
\eqref{eta2}  are shown in Fig.\ \ref{fig4}. Those pertaining to Eq.\
\eqref{3.7} (this time with $N=1,2,3,4,5$) appear in Fig.\ \ref{fig5}.

As clearly seen from these two figures, the behavior of the inverse
representations is much more regular than the one displayed in  Fig.\
\ref{fig1}. This strongly supports the notion that they are more adequate if
one wants to get a reliable value for $\eta_\infty$. Note that the results
derived from Eq.\ \eqref{3.7}, (\emph{cf.} Fig.\ \ref{fig5}) show a smooth
behavior and that the estimate  $\eta_\infty(N)$ increases \emph{slowly} with
$N$. {In the three-dimensional case, it} is not clear whether $\eta_\infty(N)$
converges to the close-packing value $\eta_{\text{max}}$ for $N\to\infty$ or to
the random close-packing value $\eta_{\text{rcp}}\simeq 0.64$. However, the
very good fit obtained with a law of the form
$\eta_\infty(N)=\eta_{\text{max}}-a e^{-bN}$ suggests that the values obtained
up to $N=5$ are not incompatible with the result
$\lim_{N\to\infty}\eta_\infty(N)=\eta_{\text{max}}$. {In the two-dimensional
case the approach of $\eta_\infty(N)$ to $\eta_\text{max}$ is much clearer than
in the three-dimensional case, the value of $\eta_\infty(5)$ being even
slightly larger than $\eta_{\text{rcp}}\simeq 0.82$. This gives further support
 to the correctness of the conjecture \eqref{18}.}

Now we turn to Fig.\ \ref{fig4}. Although the representation \eqref{3.8} leads
to one point less in the graph than Eq.\ \eqref{3.7} and does not make use of
$b_{10}$,   $\eta_\infty(4)$ is remarkably close to
$\eta_{\text{max}}${, especially in the case $d=2$}. This could be just a
coincidence and it is conceivable that, once $b_{11}$ became available and the
point corresponding to $N=5$ were included, $\eta_\infty(5)$ would turn out to
be greater than $\eta_{\text{max}}$. In fact, the fit to a law
$\eta_\infty(N)=\eta_{\text{max}}-a e^{-bN}$ is much poorer in this instance.
Further, we have found that the results for $\eta_{\infty}(N)$ obtained from
Eq.\ \eqref{3.8} are more sensitive to variations of the values of the virial
coefficients than those obtained from Eq.\ \eqref{3.7}.

In view of the above, and besides the physical reasons we alluded to before, we
find that the pressure representation seems to be the {most} reliable one for
estimating the true value of  $\eta_\infty$.  On the other hand, it seems also
clear that, strictly speaking, the knowledge of the first ten virial
coefficients is not enough to decide whether $\eta_\infty=\eta_{\text{max}}$
or $\eta_\infty=\eta_{\text{rcp}}$ {in the three-dimensional case}.

In order to look into the performance of the representation \eqref{3.7} in more
detail, we have carried out a test of robustness similar to the one made in
connection with the poles of the conventional approximants in Fig.\ \ref{fig2}.
In this case, however, rather than considering for each $N$ only an error of
$\pm 5\%$ in the values of the highest virial coefficients, we take an error of
$\pm 100\%$ in $b_{2N}$. The results of this procedure are {also} shown in
Fig.\ \ref{fig5}, where it is clear that as one increases $N$ the effect of the
error becomes less pronounced, to the point that for $N=5$ {and $d=3$} it
becomes practically unnoticeable. This means that one cannot rely on the
representation \eqref{3.7} for getting estimates of unknown virial
coefficients. On the other hand, its introduction was not made for that purpose
but rather for obtaining good estimates of $\eta_\infty$.

In this regard, albeit on a more speculative basis, for the sake of going beyond
the limit imposed by the value $N=5$ one can incorporate into the procedure
estimates of unknown higher virial coefficients $b_{11}$, $b_{12}$, \ldots.
Given the robustness of the results as discussed above, one would expect that
the effect of errors in such estimates on $\eta_\infty(N)$ for $N$ somewhat
greater than $5$ would be weak. The idea that this is indeed the case {goes
back to Eqs.\ \eqref{012D}--\eqref{04}. Let us specialize here to $d=3$. {}From
Eqs.\ \eqref{01} and \eqref{02} one obtains}
\beqa
\alpha_4&=&\frac{120.505}{1-1.30\times 10^{-3}b_9}\left(1-2.64\times
10^{-3}b_9\right.\nn
&&\left.-1.068\times 10^{-6}b_9^2+4.40\times 10^{-4}b_{10}\right),
\label{5}
\eeqa
\beqa
\beta_5&=&\frac{211.567}{1-1.30\times 10^{-3}b_9}\left(1-2.65\times
10^{-3}b_9\right.\nn
&&\left.-1.31\times 10^{-6}b_9^2+4.79\times 10^{-4}b_{10}\right)
\label{6}
\eeqa
for the coefficients of highest degree in the numerator and denominator of the
Pad\'e approximant \eqref{3.7} with $N=5$.
{Since  $b_9\sim b_{10}\sim 10^2$, it turns out that $\alpha_4$ and $\beta_5$
are hardly affected by the precise values of $b_9$ and $b_{10}$.} But, because
of a partial cancelation of terms, this influence is still weaker in the case
of the ratio $\alpha_4/\beta_5$, namely
\beqa
\eta_\infty(5)
&\simeq&0.5696\left(1+4.6\times 10^{-6}b_9+2.5\times
10^{-7}b_9^2\right.\nn
&&\left.-3.9\times 10^{-5}b_{10} \right).
\label{7}
\eeqa
When replacing the known values $b_9=85.813$ and $b_{10}=105.78$ one gets
$\eta_\infty(5)=0.5696\times (1-0.002)$, which only differs $0.2\%$ from the
value obtained by setting $b_9=b_{10}=0$.

Let us repeat the same process in the case $N=6$.{{} It can be checked that
Eqs.\ \eqref{03} and \eqref{04} yield}
\beqa
\alpha_5&=&\frac{291.651}{1-2.95\times 10^{-4}b_{11}}\left(1-5.98\times
10^{-4}b_{11}\right.\nn
&&\left.-6.03\times 10^{-8}b_{11}^2+1.11\times 10^{-4}b_{12}\right),
\label{8}
\eeqa
\beqa
\beta_6&=&\frac{482.486}{1-2.95\times 10^{-4}b_{11}}\left(1-5.99\times
10^{-4}b_{11}\right.\nn
&&\left.-6.98\times 10^{-8}b_{11}^2+1.18\times 10^{-4}b_{12}\right),
\label{9}
\eeqa
\beqa
\eta_\infty(6)
&\simeq&0.6045\left(1+5.1\times 10^{-7}b_{11}+9.4\times
10^{-9}b_{11}^2\right.\nn
&&\left.-7.1\times 10^{-6}b_{12} \right).
\label{10}
\eeqa
Inserting the estimated values $b_{11}\simeq 128$, $b_{12}\simeq153$ (see Table
\ref{tab3}) we obtain $\eta_\infty(6)=0.6045\times (1-0.0008)$, which deviates
less than $0.1\%$ from the value corresponding to $b_{11}=b_{12}=0$.
Considering the above as an illustrative example, it is not unreasonable to
expect that, as $N$ increases, the value {of  $\eta_\infty(N)$} becomes less
and less sensitive to the actual values of the virial coefficients $b_j$ with
$j\lesssim 2N$.

\begin{table}
\caption{Estimated virial coefficients\cite{CM06} $b_{11}$--$b_{16}$ for a
hard-disk fluid ($d=2$) and a hard-sphere fluid ($d=3$).
\label{tab3}}
\begin{ruledtabular}
\begin{tabular} {ccc}
$j$ &$b_j$ ($d=2$)&$b_j$ ($d=3$)\\
\hline
${11} $ & $11.15$&$128$\\
$ {12} $ & $12.08$&$153$\\
$ {13} $ & $13.03$&$182$\\
$ {14} $ & $13.93$&$215$\\
${15} $ & $14.91$&$247$\\
$ {16} $ & $15.86$&$279$\\
\end{tabular}
\end{ruledtabular}
\end{table}

\begin{figure}
\includegraphics[width=.95\columnwidth]{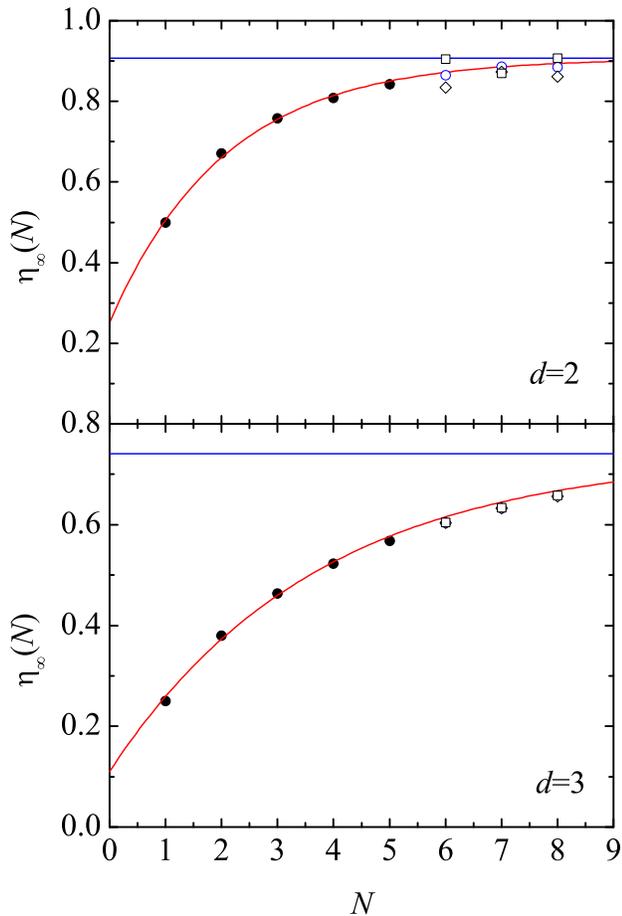}
\caption{Estimates of $\eta_\infty$ derived from Eq.\ \protect\eqref{3.7}. The
results up to $N=5$ {(filled circles)} have been derived from the known values
of the virial coefficients as given in Table \protect\ref{tab1}, while those
for $N=6$, $7$, and $8$ {(open circles) have been obtained from} the estimates
of the higher virial coefficients given in Table \protect\ref{tab3}. {The open
squares and diamonds represent the values of $\eta_\infty(N)$ when the
estimates $b_{11}$--$b_{16}$ of Table \protect\ref{tab3} are assumed to have an
increasing error between $\pm 10\%$ and $\pm 30\%$ ($d=2$) or between $\pm
20\%$ and $\pm 60\%$ ($d=3$), as explained in the text.} The horizontal lines
correspond to $\eta_{\text{max}}$. The curves correspond to the same fits as in
Fig.\ \protect\ref{fig5}.
\label{fig6}}
\end{figure}

With the previous assertion in mind, here we will go up to $N=8$ taking the
values of $b_{11}$--$b_{16}$ as estimated by Clisby and McCoy\cite{CM06} from
Pad\'e approximants. These estimates are given in Table \ref{tab3}. The
corresponding results for $\eta_\infty(N)$ are shown in Fig.\ \ref{fig6}. As
was done in Fig.\ \ref{fig5}, in order to illustrate representative outcomes,
we have also included in Fig.\ \ref{fig6} for  $N=6$, $7$, and $8$ the points
obtained when the values of $b_{11}$ and $b_{12}$ from Table \ref{tab3} {for
$d=2$ ($d=3$) have an error of $\pm 10\%$ ($\pm 20\%$), those of $b_{13}$ and
$b_{14}$  an error of $\pm 20\%$ ($\pm 40\%$), and those of $b_{15}$ and
$b_{16}$ an error of $\pm30\%$ ($\pm60\%$)}. Once more, irrespective of the
errors, one can not distinguish the three points obtained in this way with the
same $N\geq 6$ {in the case $d=3$}. Apart from this, perhaps the most
interesting feature that one can observe from Fig.\ \ref{fig6} is that the new
additions appear to continue the smooth trend obtained with the previous five
points. In fact, they fall on top of the curve that served to fit these
previous points in Fig.\ \protect\ref{fig5}. Further, the last point {of the
curve corresponding to $d=3$} is above the random close-packing value
$\eta_{\text{rcp}}\simeq 0.64$. Hence, all this evidence strongly supports the
conjecture  \eqref{18}. However, a word of caution is called for since,
irrespective of the tests that we have carried out on the sensitivity of
$\eta_\infty(N)$ to errors in the values of the virial coefficients, the
possibility that some of the estimates that we have used for the higher virial
coefficients may be quite deficient does exist. One must recall that in higher
dimensions there are negative virial coefficients\cite{CM04b} and, although it
is unlikely that for hard spheres some of the coefficients $b_{11}$ to $b_{16}$
may be negative, one cannot rule it out with certainty.

{Since the results of this section face the limitation of having been derived with the (relatively) small number of available virial coefficients, one may reasonably wonder whether our proposal would stand against a more stringent test. For instance, one could examine the model of hard hexagons on a lattice,  for which an exact solution exists and where the first twenty five virial coefficients are given by Joyce,\cite{J88} or the model of hard squares on a lattice, where the first forty two virial coefficients can be obtained from the work by Baxter \emph{et al.}\cite{BET80} While the analysis (not shown but available upon request) carried out for these systems using the inverse Pad\'e representation of the pressure similar to Eq.\ \eqref{3.7} is not incompatible with the validity of the conjecture \eqref{18}, it presents large oscillations and is flawed by unphysical results arising from some defective Pad\'e approximants. This may be due to different causes which we will discuss in connection with the hard hexagon model on a lattice. On the one hand, continuous and two-dimensional discrete models show important differences. For instance, in the hard hexagon model on a lattice there is no analytic continuation of the pressure from the low  to the high density regime. In fact, the fluid branch and the solid branch both end  and meet with a zero slope at the critical density in what appears to be a continuous phase transition from the disordered to the ordered state. In this case,  no metastable states are possible. In contrast, computer simulations have revealed both for the hard-disk and the hard-sphere fluids that, in the thermodynamic limit, the low and high density branches of the equation of state do not smoothly join one another. In the case of hard disks, it is not clear yet whether the fluid-solid transition is first order or mediated by an intermediate  hexatic phase. In the case of hard spheres,  there exists compelling evidence that these systems present a first order phase transition where a two-phase region appears with a constant pressure  as the density varies from the density of the pure fluid $\rho_f$ to the density of the pure solid $\rho_s$. The flat portion of the isotherm which connects $\rho_f$ and $\rho_s$ is the tie line. The fluid branch for $\rho>\rho_f$ is metastable but can be observed in computer simulations. Thus, while in the case of the hard hexagon model on a lattice one does know that the pressure diverges at the close-packing density in the solid branch, the question of where the analytic continuation of the fluid branch may diverge loses all meaning. On the other hand, as mentioned in Sec.\ \ref{intro}, for the hard-disk and hard-sphere fluids the existence of metastable states past the freezing point legitimately allows one to ask at which density this metastable fluid branch, when analytically continued, diverges. The present results suggest that knowledge of the virial coefficients may indeed indicate when this will occur for (continuous) hard-core fluids.}

\section{Concluding remarks}
\label{sec5}

On the basis of our results, the following conclusions and remarks can be made.
Two issues have been addressed in this paper. The first one concerns the
question of whether the known virial coefficients (presently ten both for hard
disks and hard spheres) are enough to tell us about the packing fraction
$\eta_\infty$ at which the fluid equation of state, continued and extrapolated
beyond the fluid-solid transition, has a divergence to infinity. In connection
with this issue, the determination of $\eta_\infty$ from the direct Pad\'e
approximants [\emph{cf.} Eq.\ \eqref{2.2}] seems to be not very reliable. Given
the fact that as a thermodynamic variable the pressure has a clear physical
meaning not shared by the compressibility factor, the inverse representation
\eqref{3.7} appears then as the natural candidate for such a determination, at
least when restricting to the presently available information, namely only the
ten known virial coefficients.

The second issue is related to the validity of the conjecture \eqref{18}. {In
the case of hard disks, all the results support it. On the other hand,
strictly} speaking, for the hard-sphere system our results allow us neither to
validate nor to discard this conjecture. However, {the robustness analysis
carried out for hard spheres}  strongly suggests that the conjecture is true.
Nevertheless, full confirmation must await the availability of higher virial
coefficients. In any case, we are persuaded of the usefulness of the inverse
representation  \eqref{3.7} in order to finally clarify the issue.

\begin{acknowledgments}
Two of us (M.A.G.M. and A.S.) acknowledge the financial support from the
Ministerio de  Ciencia e Innovaci\'on (Spain) through Grant No.\ FIS2010-16587
(partially financed by FEDER funds) and from the Junta de Extremadura (Spain) through Grant No.\ GR1015. The work of of M.R. and M.L.H. has been
partially supported by DGAPA-UNAM under project IN-107010-2.
\end{acknowledgments}


\begin{thebibliography}{}

\bibitem{T1885}
M. Thiesen, Ann. Phys. \textbf{24}, 467 (1885).

\bibitem{KO901}
H. Kamerlingh Onnes,  Commun. Phys. Lab. Univ. Leiden \textbf{71}
3 (1901); Proc. Koninkl. Acad. Wetensch., Amsterdam \textbf{4}, 125 (1902).

\bibitem{MM40}
J. E. Mayer and M. G. Mayer, \textit{Statistical
Mechanics} (Wiley, NY, 1940), ch.\ 13.

\bibitem{M08}
\emph{Theory and Simulation of Hard-Sphere Fluids and Related Systems},
Lectures Notes in Physics,  edited by A. Mulero (Springer, Berlin, 2008), Vol.\ \textbf{753}.

\bibitem{vdW899}
J. D. van der Waals, Proc. Koninkl. Acad. Wetensch., Amsterdam \textbf{1}, 138
(1899).

\bibitem{J896}
G. J\"{a}ger, Sitzber. Akad. Wiss. Wien Ber. Math. Natur-w. Kl. (Part 2a)
\textbf{105}, 15 (1896).


\bibitem{B896}
L. Boltzmann, Sitzber. Akad. Wiss. Wien. Ber. Math. Natur-w. Kl. (Part 2a)
\textbf{105}, 695 (1896).

\bibitem{vL899}
J. J. van Laar, Proc. Koninkl. Acad. Wetensch., Amsterdam \textbf{1}, 273
(1899).

\bibitem{B899}
L. Boltzmann, Proc. Koninkl. Acad. Wetensch., Amsterdam \textbf{1}, 398 (1899).

\bibitem{T36}
L. Tonks, Phys. Rev.  \textbf{50}, 955 (1936).

\bibitem{R64}
J. S. Rowlinson, Mol. Phys. \textbf{7}, 593 (1964).

\bibitem{H64}
P. C. Hemmer, J. Chem. Phys. \textbf{42},
1116 (1964).

\bibitem{LB82}
M. Luban and A. Baram, J. Chem. Phys. \textbf{76}, 3233 (1982).

\bibitem{BC87}
M. Baus and J. L. Colot, Phys. Rev. A \textbf{36}, 3912 (1987).

\bibitem{CM04a}
N. Clisby and B. M. McCoy, J. Stat Phys. \textbf{114}, 1343 (2004).

\bibitem{L05}
I. Lyberg, J. Stat Phys. \textbf{119}, 747 (2005).

\bibitem{MRRT53}
N. Metropolis, A. W. Rosenbluth, M. N. Rosenbluth,
and A. H. Teller, J. Chem. Phys. \textbf{21}, 1087 (1953).

\bibitem{RR54}
M. N. Rosenbluth and A. W. Rosenbluth, J. Chem. Phys. \textbf{22}, 881 (1954).

\bibitem{RH64a}F. H. Ree and W. G. Hoover, J. Chem. Phys. \textbf{40},
939 (1964).

\bibitem{RH64b}
F. H. Ree and W. G. Hoover, J. Chem. Phys. \textbf{41},
1635 (1964).

\bibitem{R65}
J. S. Rowlinson, Rep. Progr. Phys. \textbf{28}, 169 (1965).

\bibitem{RH67}
F. H. Ree and W. G. Hoover, J. Chem. Phys. \textbf{46},
4181 (1967).

\bibitem{KH68}
S. Kim and D. Henderson, Phys. Lett. A \textbf{27},
378 (1968).

\bibitem{K76}
K. W. Kratky, Physica A \textbf{85},
607 (1976).

\bibitem{K77}
K. W. Kratky, Physica A \textbf{87},
584 (1977).

\bibitem{K82a}
K. W. Kratky, J. Stat Phys. \textbf{27}, 533 (1982).

\bibitem{K82b}
K. W. Kratky, J. Stat Phys. \textbf{29}, 129 (1982).

\bibitem{vRT92}
E. J. Janse van Rensburg and G. M. Torrie, J. Phys. A: Math. Gen. \textbf{26},
943 (1992).

\bibitem{vR93}
E. J. Janse van Rensburg, J. Phys. A: Math. Gen. \textbf{26},
4805 (1993).

\bibitem{VYM02}
A. Y. Vlasov, X. M. You, and A. J. Masters, Mol. Phys. \textbf{100}, 3313
(2002).

\bibitem{CM04b}
N. Clisby and B. M. McCoy, J. Stat Phys. \textbf{114}, 1361 (2004).

\bibitem{LKM05}
S. Lab\'{\i}k, J. Kolafa, and A. Malijevsk\'y, Phys.
Rev. E \textbf{71}, 021105 (2005).

\bibitem{CM05}
N. Clisby and B. M. McCoy, Pramana \textbf{64}, 775
(2005).

\bibitem{KR06}
J. Kolafa and M. Rottner, Mol. Phys. \textbf{104}, 3435 (2006).

\bibitem{CM06}
N. Clisby and B. M. McCoy, J. Stat Phys. \textbf{122}, 15 (2006).

\bibitem{BCW08}
M. Bishop, N. Clisby, and P. A. Whitlock, J. Chem. Phys.
\textbf{128}, 034506 (2008).

\bibitem{M10}
{B. M. McCoy, \emph{Advanced Statistical Mechanics} (Oxford University Press,
Oxford, 2010).}

\bibitem{LP64}
J. L. Lebowitz and O. Penrose, J. Math. Phys. \textbf{5}, 841 (1964).

\bibitem{FPS07}
R. Fern\'andez, A. Procacci and B. Scoppola, J. Stat Phys. \textbf{128}, 1139
(2007).

\bibitem{SH09}
A. Santos and M. L\'opez de Haro,  J. Chem. Phys. \textbf{130}, 214104 (2009).

\bibitem{AFLlR84}
V. C. Aguilera-Navarro, M. Fortes, M. de Llano,  and O. Rojo, J. Chern. Phys.
\textbf{81} 1450, (1984).

\bibitem{GJ80}
D. S. Gaunt and G. S. Joyce, J. Phys. A: Math. Gen. \textbf{13},
L211 (1980).

\bibitem{J88}
{G. S. Joyce, Phil. Trans. R. Soc. London A \textbf{325}, 643 (1988).}

\bibitem{BET80}
{R. J. Baxter, I. G. Enting,  and S. K. Tsang, J. Stat. Phys. \textbf{22},
465 (1980).}

\bibitem{lattices}
See \url{http://www.math.rwth-aachen.de/~gabriele.nebe/lattices/} for a catalogue of lattices and a table with the densest packings presently known in dimensions up to 128.

\bibitem{S94}
I. C. Sanchez, J. Chem. Phys. \textbf{101},
7003 (1994).

\bibitem{W76}
{L. V. Woodcock, J. Chem. Soc. Faraday Trans. II \textbf{72}, 731 (1976).}

\bibitem{A76}
{F. C. Andrews, J. Chem. Phys. \textbf{62}, 272 (1975); \textbf{64}, 1941
(1976).}

\bibitem{BL79}
A. Baram and M. Luban, J. Phys. C \textbf{12},
L659 (1979).

\bibitem{DS82}
{J. A. Devore and E. Schneider, J. Chem. Phys. \textbf{77}, 1067 (1982).}

\bibitem{AFLlPRR83}
V. C. Aguilera-Navarro, M. Fortes, M. de Llano, A. Plastino, and O. Rojo, J.
Stat Phys. \textbf{32}, 95 (1983).

\bibitem{HvD84}
{B. R. Hoste and W. van Dael, J. Chem. Soc. Faraday Trans. II \textbf{80}, 477
(1984).}

\bibitem{GW88}
{J. I. Goldman and J. A. White, J. Chem. Phys. \textbf{89}, 6403 (1988).}

\bibitem{SHY95}
A. Santos, M. L\'opez de Haro, and S. B. Yuste,  J. Chem. Phys. \textbf{103},
4622 (1995).

\bibitem{WKV96}
{W. Wang, M. K. Khoshkbarchi, and J. H. Vera, Fluid Phase Equil. \textbf{115},
25 (1996).}

\bibitem{KV97}
{M. K. Khoshkbarchi and  J. H. Vera, Fluid Phase Equil. \textbf{130}, 189
(1997).}

\bibitem{NAM00}
{Kh. Nasrifar, Sh. Ayatollahi, and M. Moshfeghian, Can. J. Chem. Eng.
\textbf{78},
1111 (2000).}

\bibitem{GV01}
{C. Ghotbi and J. H. Vera, Can. J. Chem. Eng. \textbf{79}, 678 (2001).}

\bibitem{W02}
{X. Z. Wang, Phys. Rev. E \textbf{66}, 31203 (2002).}

\bibitem{PV05}
{I. Polishuk and J. H. Vera, J. Phys. Chem. B \textbf{109}, 5977 (2005).}

\bibitem{MMD06}
{M. Miandehy, H. Modarress, and M. R. Dehghani, Fluid Phase Equil. \textbf{239},
91
(2006).}

\bibitem{K92}
{D. I. Korteweg,  Nature \textbf{45}, 277 (1892).}

\bibitem{B98}
L. Boltzmann, \emph{Vorlesungen \"uber Gastheorie} (Barth,
Leipzig, 1898), Vol.\ 2, Chap. 5 [\emph{Lectures on Gas Theory}, Part 2, translated by S. G. Brush (University of California Press, Berkeley, CA, 1964, 2006), Chap. 5 (in English)].



\bibitem{F72}
{E. J. Le Fevre, Nature (London) Phys. Sci. \textbf{235}, 20 (1972).}

\bibitem{AFLlPR82}
{V. C. Aguilera-Navarro, M. Fortes, M. de Llano, and A. Plastino, J. Chem.
Phys. \textbf{76}, 749 (1982).}

\bibitem{MA86}
{D. Ma and G. Ahmadi, J. Chem. Phys. \textbf{84}, 3449 (1986).}

\bibitem{SSM88}
{Y. Song, R. M. Stratt, and E. A. Mason, J. Chem. Phys. \textbf{88}, 1126
(1988).}

\bibitem{H97}
{E. Z. Hamad, Ind. Eng. Chem. Res. \textbf{36}, 4385 (1997).}

\bibitem{B83}
{J. G. Berryman, Phys. Rev. A \textbf{27}, 1053 (1983).}

\bibitem{MGPC08}
A. Mulero, C. A. Gal\'an, M. I. Parra, and F. Cuadros, in \emph{Theory and Simulation of Hard-Sphere Fluids and Related Systems},
Lectures Notes in Physics,  edited by A. Mulero (Springer, Berlin, 2008), Vol.\ \textbf{753}, pp.\ 37--109.


\bibitem{SL09}
 I. C. Sanchez and J. S. Lee, J. Phys. Chem. B. \textbf{113}, 15572 (2009).




\bibitem{BO78}
 C. Bender and S. Orszag, \emph{Advanced Mathematical Methods for Scientists and
Engineers} (McGraw-Hill, NY, 1978), ch.\ 8.



\bibitem{GB08}
A. O. Guerrero and A. Bassi, J. Chem. Phys. \textbf{129}, 044509 (2008).

\bibitem{HY09}
J. Hu and Y. Yu, Phys. Chem. Chem. Phys.\textbf{ 11}, 9382 (2009).

\bibitem{HY00}
{E. Z. Hamad and G. O. Yahaya, Fluid Phase Equilib. \textbf{168} 59, (2000).}

\bibitem{SZK53}
Z. W. Salsburg, R. W. Zwanzig, and J. G. Kirkwood, J. Chem. Phys. \textbf{21},
1098 (1953).

\bibitem{CD98}
D. S. Corti and P. G. Debenedetti, Phys. Rev. E \textbf{57}, 4211 (1998).

\bibitem{HC04}
M. Heying and D. S. Corti, Fluid Phase Equilib. \textbf{220}, 85 (2004).

\bibitem{FGMS09}
R. Fantoni, A. Giacometti, Al. Malijevsk\'y, and A. Santos,  J. Chem. Phys.
\textbf{131}, 124106 (2009).

\bibitem{BNS09}
A. Ben-Naim and A. Santos,  J. Chem. Phys. \textbf{131}, 164512 (2009).




\end{thebibliography}
\end{document}